# The relation between Eigenfactor, audience factor, and influence weight

Ludo Waltman and Nees Jan van Eck

Centre for Science and Technology Studies, Leiden University, The Netherlands
{waltmanlr, ecknjpvan}@cwts.leidenuniv.nl

We present a theoretical and empirical analysis of a number of bibliometric indicators of journal performance. We focus on three indicators in particular, namely the Eigenfactor indicator, the audience factor, and the influence weight indicator. Our main finding is that the last two indicators can be regarded as a kind of special cases of the first indicator. We also find that the three indicators can be nicely characterized in terms of two properties. We refer to these properties as the property of insensitivity to field differences and the property of insensitivity to insignificant journals. The empirical results that we present illustrate our theoretical findings. We also show empirically that the differences between various indicators of journal performance are quite substantial.

## Introduction

The impact factor (Garfield, 1972, 2006) is without doubt the most commonly used bibliometric indicator of the performance of scientific journals. Various alternatives to the impact factor have been proposed in the literature. These alternatives include indicators based on cited-side normalization (e.g., Van Leeuwen & Moed, 2002), indicators based on citing-side normalization (Moed, in press; Zitt & Small, 2008), indicators based on the *h*-index (e.g., Braun, Glänzel, & Schubert, 2006), and indicators based on recursive citation weighting. Indicators based on recursive citation weighting were first proposed by Pinski and Narin (1976; see also Geller, 1978), and they have been popular in the field of economics (Kalaitzidakis, Mamuneas, & Stengos, 2003; Kodrzycki & Yu, 2006; Laband & Piette, 1994; Liebowitz & Palmer, 1984; Palacios-Huerta & Volij, 2004). The successful PageRank algorithm of the Google search engine (Brin & Page, 1998; Page, Brin, Motwani, & Winograd, 1998; see also Langville & Meyer, 2006) has caused a renewed interest in recursive indicators of journal performance. Three PageRank-inspired indicators that have been recently introduced are the weighted PageRank indicator (Bollen, Rodriguez, & Van de Sompel, 2006; Dellavalle, Schilling, Rodriguez, Van de Sompel, & Bollen, 2007), the Eigenfactor indicator (Bergstrom, 2007; West, Bergstrom, & Bergstrom, in press), and the SCImago Journal Rank indicator (González-Pereira, Guerrero-Bote, & Moya-Anegón, 2009).

In this paper, we point out the relation between three indicators of journal performance, namely the audience factor (Zitt & Small, 2008), the influence weight indicator (Pinski & Narin, 1976), and the Eigenfactor indicator. The audience factor is based on citing-side normalization, while the other two indicators are based on recursive citation weighting. Unlike the audience factor and the influence weight indicator, the Eigenfactor indicator is a parameterized indicator. Hence, the behavior of the Eigenfactor indicator depends on the choice of a parameter. Our main finding is that the audience factor and the influence weight indicator can be regarded as a kind of special cases of the Eigenfactor indicator. Related to this, we show how the three indicators can be characterized in terms of two properties that we introduce. We refer to these properties as the property of insensitivity to field differences and the property of insensitivity to insignificant journals. Interestingly, it turns out that the parameter of the Eigenfactor indicator can be used to make a trade-off between the two properties. In addition to a theoretical analysis of the audience factor, the influence weight indicator, and the Eigenfactor indicator, we also report some results of an empirical analysis of these indicators.

This paper is organized as follows. First, we discuss our indicators of interest and we point out how these indicators are mathematically related to each other. Next, we study the indicators empirically. Finally, we briefly discuss some other related indicators and we summarize our conclusions. Some technical details are elaborated in an appendix.

## Indicators

In this section, we discuss the indicators that we study in this paper. We use the following mathematical notation. Let there be $n$ journals, denoted by 1, …, $n$. Let $T_1$ and $T_2$ denote two time periods, where period $T_1$ precedes period $T_2$. (The two periods may overlap or coincide.) We are interested in measuring the performance of journals 1, …, $n$ based on citations from articles published in period $T_2$ to articles published in period $T_1$. Let $a_{i1}$ and $a_{i2}$ denote the number of articles published in journal $i$ in, respectively, periods $T_1$ and $T_2$, and let $c_{ij}$ denote the number of citations from articles published in journal $i$ in period $T_2$ to articles published in journal $j$ in period $T_1$. We define $s_i$ as

$$s_i = \sum_j c_{ij} . \tag{1}$$

Hence, $s_i$ denotes the total number of citations from articles published in journal $i$ in period $T_2$ to articles published in journals 1, …, $n$ in period $T_1$.

Using the above mathematical notation, we now discuss our indicators of interest. We focus on the essential characteristics of the indicators. We ignore practical issues such as the document types (e.g., articles, letters, and reviews) that are taken into account, the length of the time window within which citations are counted, and the way in which self citations are handled. For our present purposes, issues such as these are not important.

*Impact factor*

Although the impact factor is not our main interest in this paper, we include it for completeness. The impact factor is defined as the average number of citations that a journal has received per article (Garfield, 1972, 2006). Hence, the impact factor of journal $i$ can be written as

$$\mathrm{IF}_i = \frac{\sum_j c_{ji}}{a_{i1}} . \tag{2}$$

The impact factor is a very simple indicator. It is well known that in some fields articles are on average cited much more frequently than in other fields. The impact factor does not correct for such differences among fields. Because of this, impact factors of journals in different fields should not be directly compared with each other.

*Audience factor*

The audience factor is a recent proposal of Zitt and Small (2008). The audience factor is similar to the impact factor except that citations are weighted based on the journal from which they originate. The larger a journal's average number of references per article, the lower the weight of a citation originating from the journal. The audience factor of journal $i$ is defined as

$$\mathrm{AF}_i = \frac{1}{a_{i1}} \sum_j \frac{m_\mathrm{S}}{m_j} c_{ji} , \tag{3}$$

where $m_j$ and $m_\mathrm{S}$ are given by

$$m_j = \frac{s_j}{a_{j2}} , \tag{4}$$

$$m_\mathrm{S} = \frac{\sum_j s_j}{\sum_j a_{j2}} , \tag{5}$$

that is, $m_j$ denotes journal $j$'s average number of references per article and $m_\mathrm{S}$ denotes the average number of references per article for all journals taken together. Notice that in the definitions of $m_j$ and $m_\mathrm{S}$ only references to articles published in journals 1, …, $n$ in period $T_1$ are taken into account. These references are called active references by Zitt and Small. All non-active references are ignored.

By assigning weights to citations, the audience factor aims to correct for differences among fields. Unlike indicators based on cited-side normalization (e.g., Van Leeuwen & Moed, 2002), which also aim to correct for field differences, the audience factor has the advantage that it does not rely on an externally imposed field classification. In Appendix A, we introduce the property of *insensitivity to field differences*. This property provides a formal definition of the idea of correcting for field differences. Informally, the property of insensitivity to field differences has the following interpretation. Suppose that we have two equally-sized fields and that each journal gives away only a small amount of citations to journals that are not in its own field. We then say that an indicator is insensitive to field differences if the average value of the indicator for one field deviates from the average value of the indicator for the other field only by a small amount. In the case of two fields without any between-fields citation traffic, the property of insensitivity to field differences requires that the average value of an indicator is the same for both fields. We show in appendix A that under a relatively mild assumption the audience factor has the property of insensitivity to field differences.

*Influence weight*

The influence weight indicator was proposed by Pinski and Narin (1976). The influence weights of journals 1, …, $n$, denoted by $\mathrm{IW}_1$, …, $\mathrm{IW}_n$, are obtained by solving the following system of linear equations:[1]

[1] This system of linear equations has a unique solution if the journal citation matrix $\mathbf{C} = [c_{ij}]$ is irreducible. In other words, the system of linear equations has a unique solution if in the journal citation graph there exists for any two journals $i$ and $j$ a path from $i$ to $j$ and a path from $j$ to $i$. We note that for

$$\mathrm{IW}_i = \frac{\sum_j \mathrm{IW}_j c_{ji}}{s_i} \qquad \text{for } i = 1, \ldots, n \tag{6}$$

$$\frac{\sum_i \mathrm{IW}_i s_i}{\sum_i s_i} = 1 \,. \tag{7}$$

Unlike the impact factor and the audience factor, the influence weight indicator is a measure of a journal's average performance per reference rather than of its average performance per article. Based on the influence weight of journal $i$, a measure of journal $i$'s average performance per article can be obtained by

$$\mathrm{IPP}_i = \frac{\mathrm{IW}_i s_i}{a_{i1}} \,. \tag{8}$$

Following Pinski and Narin (1976), we refer to the indicator in Equation 8 as the influence per publication indicator. Theoretical studies of the influence weight indicator and the influence per publication indicator can be found in papers by Geller (1978), Palacios-Huerta and Volij (2004), and Serrano (2004). In the last two papers, the influence per publication indicator is referred to as the invariant method.

In Appendix A, we show that the influence per publication indicator does not have the property of insensitivity to field differences. However, the influence per publication indicator does have another interesting property, referred to as the property of *insensitivity to insignificant journals*. To see this, consider the following example. There are $n$ = 8 journals. Each journal publishes 100 articles in each time period. Hence, $a_{i1} = a_{i2} = 100$ for $i = 1, \ldots, n$. The journal citation matrix $\mathbf{C} = [c_{ij}]$ is shown in Table 1. Based on this matrix, two fields can be distinguished. One field consists of journals 1, 2, 3, and 4. The other field consists of journals 5, 6, 7, and 8. A distinction can also be made between frequently cited journals and infrequently cited journals. Journals 1, 2, 5, and 6 are frequently cited, while journals 3, 4, 7, and 8 are infrequently cited. In practice, it is almost impossible to have publication and citation data for all infrequently cited journals in a field. This is because the coverage of infrequently cited journals in bibliographic databases such as Web of Science and Scopus is far from complete. Some infrequently cited journals are covered by these databases, but many others are not. To examine the consequences of incomplete coverage of infrequently cited journals, we look at two scenarios, scenario 1 and scenario 2. In scenario 1, journals 1, …, 8 are all covered by the bibliographic database that we use. In scenario 2, journals 1, …, 7 are covered while journal 8 is not. For both scenarios, influence per publication scores calculated using Equations 6, 7, and 8 are reported in Table 2. As can be seen in the table, the influence per publication scores of journals 1, …, 7 are very similar in the two scenarios. This demonstrates that the influence per publication indicator is rather insensitive to incomplete coverage of infrequently cited journals. We therefore say that the influence per publication indicator has the property of insensitivity to insignificant journals.

---

computational reasons it is convenient if the system of linear equations is not only irreducible but also aperiodic.

TABLE 1. Journal citation matrix. Rows correspond with citing journals. Columns correspond with cited journals.

| Journal | 1 | 2 | 3 | 4 | 5 | 6 | 7 | 8 |
|---|---|---|---|---|---|---|---|---|
| 1 | 1000 | 1000 | 10 | 10 | 100 | 100 | 1 | 1 |
| 2 | 1000 | 1000 | 10 | 10 | 100 | 100 | 1 | 1 |
| 3 | 1000 | 1000 | 10 | 10 | 100 | 100 | 1 | 1 |
| 4 | 1000 | 1000 | 10 | 10 | 100 | 100 | 1 | 1 |
| 5 | 100 | 100 | 1 | 1 | 1000 | 1000 | 10 | 10 |
| 6 | 100 | 100 | 1 | 1 | 1000 | 1000 | 10 | 10 |
| 7 | 100 | 100 | 1 | 1 | 1000 | 1000 | 10 | 10 |
| 8 | 100 | 100 | 1 | 1 | 1000 | 1000 | 10 | 10 |

TABLE 2. Journals' influence per publication scores and audience factors.

| Journal | 1 | 2 | 3 | 4 | 5 | 6 | 7 | 8 |
|---|---|---|---|---|---|---|---|---|
| IPP (scenario 1) | 5.500 | 5.500 | 0.055 | 0.055 | 5.500 | 5.500 | 0.055 | 0.055 |
| IPP (scenario 2) | 5.513 | 5.513 | 0.055 | 0.055 | 5.490 | 5.490 | 0.055 | |
| AF (scenario 1) | 44.000 | 44.000 | 0.440 | 0.440 | 44.000 | 44.000 | 0.440 | 0.440 |
| AF (scenario 2) | 42.938 | 42.938 | 0.429 | 0.429 | 34.063 | 34.063 | 0.341 | |

What is the relevance of the property of insensitivity to insignificant journals? This can be seen as follows. Suppose that instead of the influence per publication indicator the audience factor is used in the above example. For both scenario 1 and scenario 2, audience factors calculated using Equations 3, 4, and 5 are reported in Table 2. Comparing the two scenarios, it is clear that the audience factor does not have the property of insensitivity to insignificant journals. Due to the non-coverage of journal 8 in scenario 2, journals 5, 6, and 7 have substantially lower audience factors in this scenario than in scenario 1. Journals 1, 2, 3, and 4 have only marginally lower audience factors. Hence, the non-coverage of journal 8 in scenario 2 causes a substantial decrease of the audience factors of journals 5, 6, and 7 relative to the audience factors of journals 1, 2, 3, and 4. The results reported in Table 2 demonstrate that, when using an indicator that does not have the property of insensitivity to insignificant journals, the score of a journal in a certain field may strongly depend on the number of infrequently cited journals in the same field that are covered by the bibliographic database that one uses. This sensitivity to infrequently cited journals may be problematic when comparing scores of journals in different fields. If the bibliographic database that one uses covers relatively more infrequently cited journals in one field than in another, journals in the former field have an advantage over journals in the latter field.

We have now introduced two properties that bibliometric indicators of journal performance may or may not have, namely the property of insensitivity to field differences and the property of insensitivity to insignificant journals. It is important to note that these two properties rule out each other, that is, an indicator cannot have both properties. The following example shows this. Suppose that an infrequently cited journal is added to the bibliographic database that one uses. The property of insensitivity to insignificant journals then requires that, because the newly added journal is infrequently cited, the scores of all other journals remain more or less unchanged. The property of insensitivity to field differences, on the other hand, requires that the average score of the journals in a field remains unchanged. Hence, in the field to which the newly added journal belongs, the scores of all other journals must increase somewhat (otherwise the newly added journal would cause a decrease of the average score of the journals in the field). Based on this example, it is clear that insensitivity to field differences and insensitivity to insignificant journals are conflicting properties that cannot be satisfied both at the same time.

*Eigenfactor*

The Eigenfactor indicator (Bergstrom, 2007; West et al., in press) is a recently proposed indicator of journal performance. The indicator belongs to the family of PageRank-inspired indicators. Eigenfactor scores of a large number of journals, calculated based on Web of Science data, are available at www.eigenfactor.org. Eigenfactor scores can also be found in the Journal Citation Reports of Thomson Reuters. Various properties of the Eigenfactor indicator are discussed by Franceschet (in press-b). Below, we focus on the essential characteristics of the Eigenfactor indicator. We ignore the way in which the Eigenfactor indicator handles journal self citations and so-called dangling nodes.

The Eigenfactor indicator is a parameterized indicator. Let $\alpha \in [0, 1]$ denote the parameter of the Eigenfactor indicator. The parameter is similar to what is often referred to as the damping factor parameter in the PageRank literature. By default, the Eigenfactor indicator uses $\alpha$ equal to 0.85. This is also the default value of the damping factor parameter in the PageRank algorithm. Eigenfactor scores are calculated as follows (West, Althouse, Rosvall, Bergstrom, & Bergstrom, 2008; West & Bergstrom, 2008; for an intuitive description, see West et al., in press). For each journal $i$, a value $p_i$ is obtained by solving the following system of linear equations:[2]

$$p_i = \alpha \sum_j \frac{p_j c_{ji}}{s_j} + (1-\alpha) \frac{a_{i1}}{\sum_j a_{j1}} \qquad \text{for } i = 1, \ldots, n \tag{9}$$

$$\sum_i p_i = 1. \tag{10}$$

Using the values $p_1, \ldots, p_n$, the Eigenfactor score of journal $i$ is given by

$$\mathrm{EF}_i(\alpha) = 100 \sum_j \frac{p_j c_{ji}}{s_j}. \tag{11}$$

Unlike for example the impact factor, the Eigenfactor indicator is a measure of a journal’s total performance rather than of its average performance per article. Hence, the Eigenfactor indicator is size dependent. Other things equal, a journal that publishes twice as many articles has a twice as high Eigenfactor score. Based on the Eigenfactor score of journal $i$, a measure of journal $i$’s average performance per article can be obtained by

$$\mathrm{AI}_i(\alpha) = \frac{\mathrm{EF}_i(\alpha)}{100\, a_{i1}}. \tag{12}$$

[2] For $\alpha < 1$, this system of linear equations always has a unique solution. For $\alpha = 1$, the system of linear equations has a unique solution if the journal citation matrix $\mathbf{C} = [c_{ij}]$ is irreducible.

The indicator in Equation 12 is referred to as the article influence indicator.

The properties of the Eigenfactor indicator and the article influence indicator depend on the parameter $\alpha$. We study this dependence in the next section.

## Relation of Eigenfactor with audience factor and influence weight

In the previous section, formal mathematical definitions of the audience factor, the influence weight indicator, and the Eigenfactor indicator were provided. Using these definitions, it is relatively easy to see the relation between the three indicators. However, we do not compare the indicators directly with each other. This is because the indicators are normalized in different ways. That is, the audience factor is a measure of a journal's average performance per article, the influence weight indicator is a measure of a journal's average performance per reference, and the Eigenfactor indicator is a measure of a journal's total performance. In this paper, we focus on measuring a journal's average performance per article. Hence, instead of the audience factor, the influence weight indicator, and the Eigenfactor indicator, we compare the audience factor, the influence per publication indicator, and the article influence indicator. As discussed in the previous section, these three indicators all measure a journal's average performance per article.

We first consider the relation between the article influence indicator and the audience factor. We assume that the number of articles that journals publish either remains stable over time or increases or decreases by the same percentage for all journals. Under this assumption, it turns out that, if the Eigenfactor parameter $\alpha$ equals 0, the article influence score of a journal is proportional to the audience factor of a journal.[3] Hence, under the above assumption, the audience factor can be regarded as a special case of the article influence indicator. This result is stated formally in the following theorem.

**Theorem 1.** Let the number of articles published in a journal in period $T_2$ be proportional to the number of articles published in a journal in period $T_1$, that is, let $a_{i2}$ be proportional to $a_{i1}$. Furthermore, let the Eigenfactor parameter $\alpha$ be equal to 0. The article influence score of a journal is then proportional to the audience factor of a journal, that is, $\mathrm{AI}_i(\alpha)$ is proportional to $\mathrm{AF}_i$.

**Proof.** Let $\alpha = 0$. It then follows from Equations 9, 11, and 12 that

$$\mathrm{AI}_i(\alpha) \propto \frac{1}{a_{i1}} \sum_j \frac{a_{j1}}{s_j} c_{ji} . \tag{13}$$

It follows from Equations 3 and 4 that

$$\mathrm{AF}_i \propto \frac{1}{a_{i1}} \sum_j \frac{a_{j2}}{s_j} c_{ji} . \tag{14}$$

Hence, if $a_{i2} \propto a_{i1}$, then $\mathrm{AI}_i(\alpha) \propto \mathrm{AF}_i$. This completes the proof of the theorem.

---

[3] Two indicators are proportional if they differ by at most a multiplicative constant. For practical purposes, indicators that are proportional can be regarded as identical.

We now consider the relation between the article influence indicator and the influence per publication indicator. It turns out that, if the Eigenfactor parameter $\alpha$ equals 1, the article influence score of a journal is proportional to the influence per publication score of a journal. Hence, the influence per publication indicator can be regarded as a special case of the article influence indicator. This result is stated formally in the following theorem.[4]

**Theorem 2.** Let the Eigenfactor parameter $\alpha$ be equal to 1. The article influence score of a journal is then proportional to the influence per publication score of a journal, that is, $\mathrm{AI}_i(\alpha)$ is proportional to $\mathrm{IPP}_i$.

**Proof.** Let $\alpha = 1$. It then follows from Equation 9 that

$$p_i = \sum_j \frac{p_j c_{ji}}{s_j}. \tag{15}$$

Equations 11 and 12 then imply that

$$\mathrm{AI}_i(\alpha) = \frac{p_i}{a_{i1}}. \tag{16}$$

Let $q_i = \mathrm{IW}_i s_i$. Equations 6, 7, and 8 can then be rewritten as, respectively,

$$q_i = \sum_j \frac{q_j c_{ji}}{s_j}, \tag{17}$$

$$\frac{\sum_i q_i}{\sum_i s_i} = 1, \tag{18}$$

$$\mathrm{IPP}_i = \frac{q_i}{a_{i1}}. \tag{19}$$

Comparing Equations 10 and 15 to Equations 17 and 18, it is clear that $p_i \propto q_i$. It then follows from Equations 16 and 19 that $\mathrm{AI}_i(\alpha) \propto \mathrm{IPP}_i$. This completes the proof of the theorem.

It follows from Theorems 1 and 2 that the article influence indicator can be regarded as a kind of interpolation between the audience factor and the influence per publication indicator. The closer the Eigenfactor parameter $\alpha$ is set to 0, the more the article influence indicator behaves like the audience factor. The closer the Eigenfactor parameter $\alpha$ is set to 1, the more the article influence indicator behaves like the influence per publication indicator. We know from the previous section that the audience factor and the influence per publication indicator have more or less opposite properties. Under a relatively mild assumption, the audience factor has the property of insensitivity to field differences. The audience factor does not have the property of insensitivity to insignificant journals. The influence per publication indicator does have the property of insensitivity to insignificant journals but does not have the property of insensitivity to field differences. It is now clear that the article influence indicator allows one to make a trade-off between the properties of insensitivity to field differences and insensitivity to insignificant journals. Setting the Eigenfactor parameter $\alpha$ close to 0 gives more weight to the property of insensitivity to field differences. Setting the Eigenfactor parameter $\alpha$ close to 1 gives more weight to the property of insensitivity to insignificant journals.

[4] Geller (1978) pointed out that the indicators proposed by Pinski and Narin (1976) can be interpreted in terms of Markov chain theory. This is the main insight needed to see the relation between the article influence indicator (as well as other PageRank-inspired indicators) and the influence per publication indicator.

## Empirical analysis

In the previous two sections, indicators of journal performance were studied theoretically. We now turn to the empirical analysis of journal performance indicators. Like in the previous section, we only consider indicators that measure a journal's average performance per article. In addition to the audience factor, the influence per publication indicator, and the article influence indicator, we also take into account the impact factor. We pay special attention to the effect of the Eigenfactor parameter $\alpha$ on the behavior of the article influence indicator.[5] For other papers in which PageRank-inspired indicators of journal performance are studied empirically, we refer to Bollen et al. (2006), Bollen, Van de Sompel, Hagberg, and Chute (2009), Davis (2008), Falagas, Kouranos, Arencibia-Jorge, and Karageorgopoulos (2008), Fersht (2009), Franceschet (2010, in press-a), Leydesdorff (2009), López-Illescas, de Moya-Anegón, and Moed (2008), and West et al. (2009, in press).[6]

Our empirical analysis is based on the Web of Science database. Only the sciences and the social sciences are considered. The arts and humanities are not taken into account. We first collected all citations from articles published in 2008 to articles published between 2003 and 2007.[7] We then selected all journals that have at least one incoming citation in each year between 2003 and 2007 and at least one outgoing citation in 2008.[8] In this way, we obtained a set of 6708 journals. For each of these journals, we calculated six performance measures, namely the impact factor, the audience factor, and the article influence score for four different values (0, 0.5, 0.85, and 1) of the Eigenfactor parameter $\alpha$. We did not calculate the influence per publication score of a journal. This is because, according to Theorem 2, influence per publication scores are perfectly correlated with article influence scores calculated for $\alpha$ equal to 1. In order to keep the analysis as transparent as possible, we calculated all performance measures exactly according to the mathematical specification provided earlier in this paper. This means that we sometimes deviated slightly from the way in which indicators were originally proposed. For example, in the case of the audience factor, we did not impose any restrictions on the weight of a citation (unlike Zitt & Small, 2008, p. 1859), and in the case of the article influence indicator, we did not ignore journal self citations (unlike West et al., 2008, p. 1).

---

[5] This is similar to the work of Ding, Yan, Frazho, and Caverlee (2009). Notice, however, that Ding et al. study authors rather than journals and that they focus on measuring total performance rather than average performance per article.

[6] Davis (2008) and Fersht (2009) compare indicators that measure a journal's total performance. West et al. (2009) criticize this approach and explain why it is better to compare indicators that measure a journal's average performance per article. Bollen et al. (2006), Davis (2008), and Franceschet (2010) compare indicators of total performance with indicators of average performance per article. It is not exactly clear to us how such comparisons should be interpreted.

[7] We only took into account the document types article and review.

[8] In addition, we required the journal citation matrix to be irreducible. This also led to the exclusion of some journals.

In Table 3, we report the Pearson and Spearman correlations between the different indicators of journal performance. The correlation between the audience factor and the article influence indicator for $\alpha$ equal to 0 is of special interest. It follows from Theorem 1 that this correlation equals 1 under the assumption that the number of articles that journals publish either remains stable over time or increases or decreases by the same percentage for all journals. Of course, in practice this assumption does not hold exactly. However, as can be seen in Table 3, our empirical results still indicate a very high correlation between the audience factor and the article influence indicator for $\alpha$ equal to 0. This correlation is also clearly visible in Figure 1, in which the empirical relation between the two indicators is shown. Based on Table 3 and Figure 1, we conclude that for most practical purposes the two indicators can be regarded as identical.

TABLE 3. Correlations between six indicators of journal performance. Pearson correlations are reported in the lower left part of the table. Spearman correlations are reported in the upper right part.

| | IF | AF | AI(0.00) | AI(0.50) | AI(0.85) | AI(1.00) |
|---|---|---|---|---|---|---|
| IF | | 0.79 | 0.79 | 0.82 | 0.87 | 0.93 |
| AF | 0.87 | | 0.98 | 0.98 | 0.92 | 0.77 |
| AI(0.00) | 0.88 | 0.99 | | 0.99 | 0.93 | 0.77 |
| AI(0.50) | 0.92 | 0.98 | 0.99 | | 0.97 | 0.84 |
| AI(0.85) | 0.93 | 0.90 | 0.91 | 0.97 | | 0.93 |
| AI(1.00) | 0.91 | 0.76 | 0.79 | 0.87 | 0.96 | |

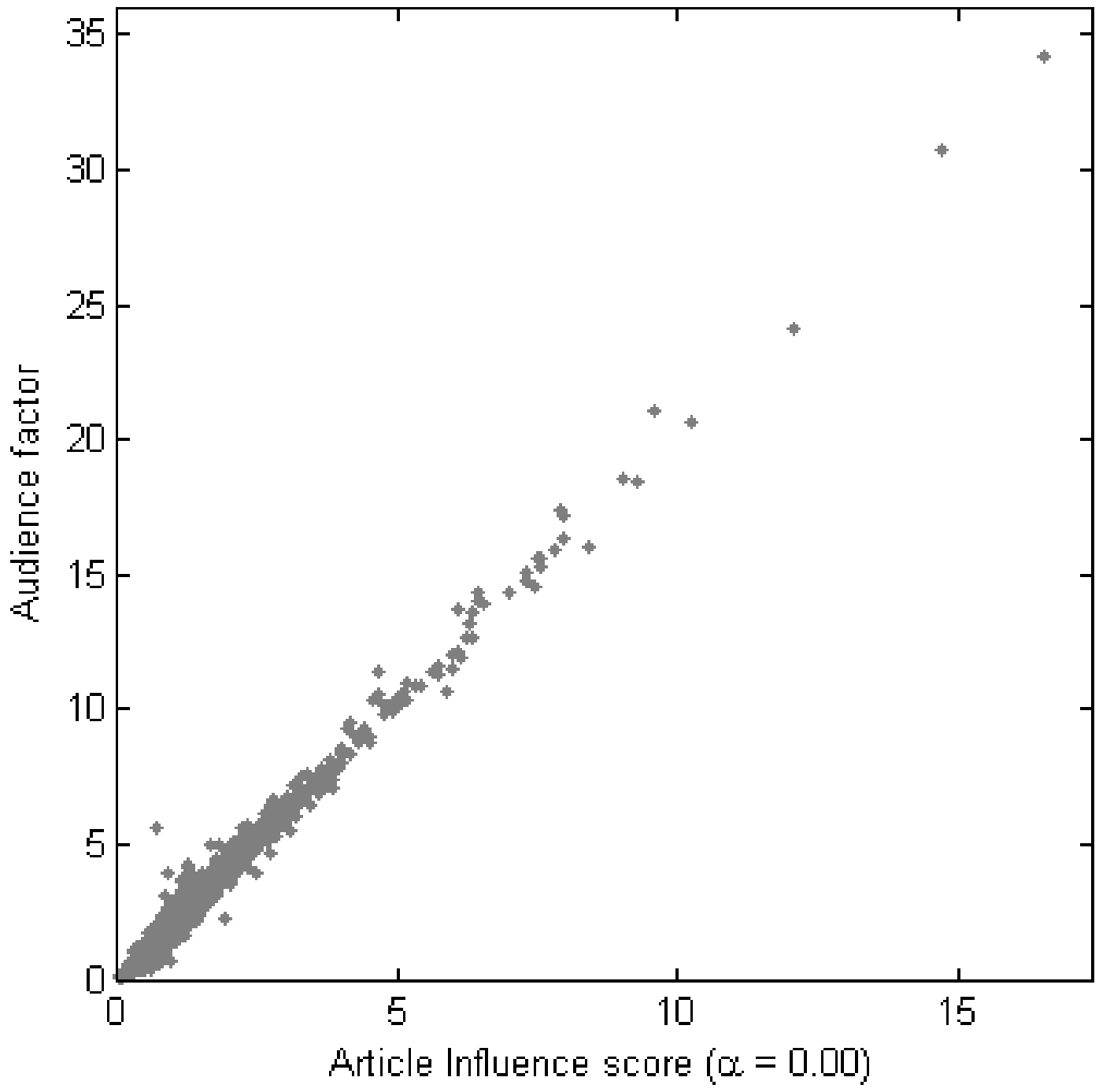

FIG. 1. Relation between the article influence indicator for $\alpha$ equal to 0 and the audience factor.

Another thing to note in Table 3 is the very high correlation between the article influence indicator for $\alpha$ equal to 0 and the article influence indicator for $\alpha$ equal to 0.5. This correlation is much closer to 1 than the correlation between the article influence indicator for $\alpha$ equal to 0.5 and the article influence indicator for $\alpha$ equal to 1. Hence, the results presented in Table 3 indicate that for higher values of $\alpha$ the article influence indicator is more sensitive to changes in $\alpha$ than for lower values of $\alpha$. For a mathematical explanation for this observation, we refer to Langville and Meyer (2006, Section 6.1).

In Figure 2, we show the empirical relations between four indicators, namely the impact factor and the article influence indicator for $\alpha$ equal to 0, 0.85, and 1. (Relations for the audience factor and for the article influence indicator for $\alpha$ equal to 0.5 are not shown. This is because these indicators are both strongly correlated with the article influence indicator for $\alpha$ equal to 0.) Although the correlations reported in Table 3 are all above 0.75, it is clear from Figure 2 that the relations between most indicators are not particularly strong. This can be seen even better by excluding the journals with the highest scores, as we do in Figure 3. When high scoring journals are excluded, relations between indicators can be rather weak.

For each of the four indicators considered in Figure 2, we list in Table 4 in Appendix B the 20 best performing journals. The results shown in Table 4 make clear that for high performing journals there can also be substantial differences between indicators. Comparing for example the article influence indicator for $\alpha$ equal to 0 and the article influence indicator for $\alpha$ equal to 1, there turn out to be only two journals that are in the top 10 for both indicators.

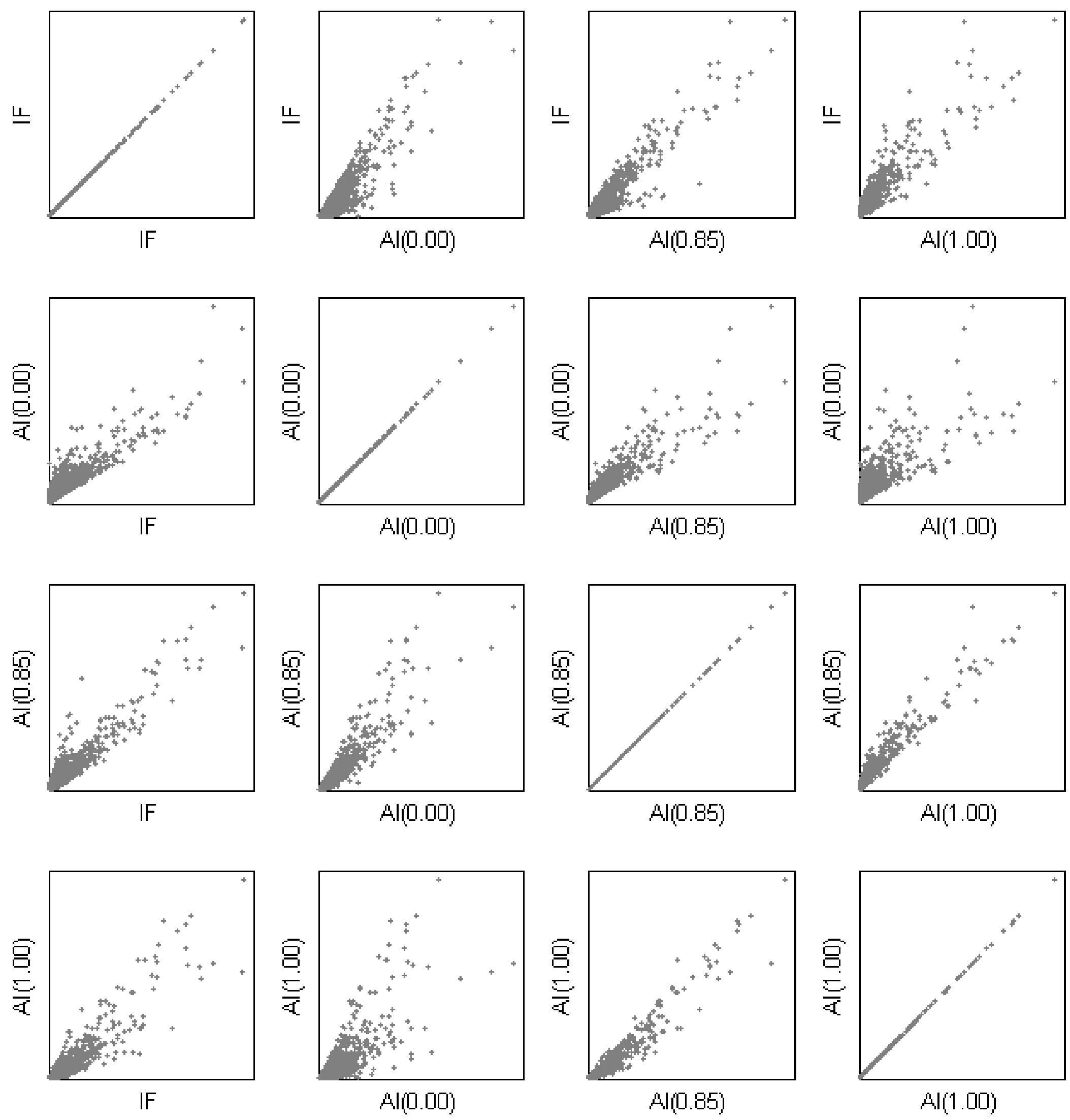

FIG. 2. Relations between four indicators of journal performance.

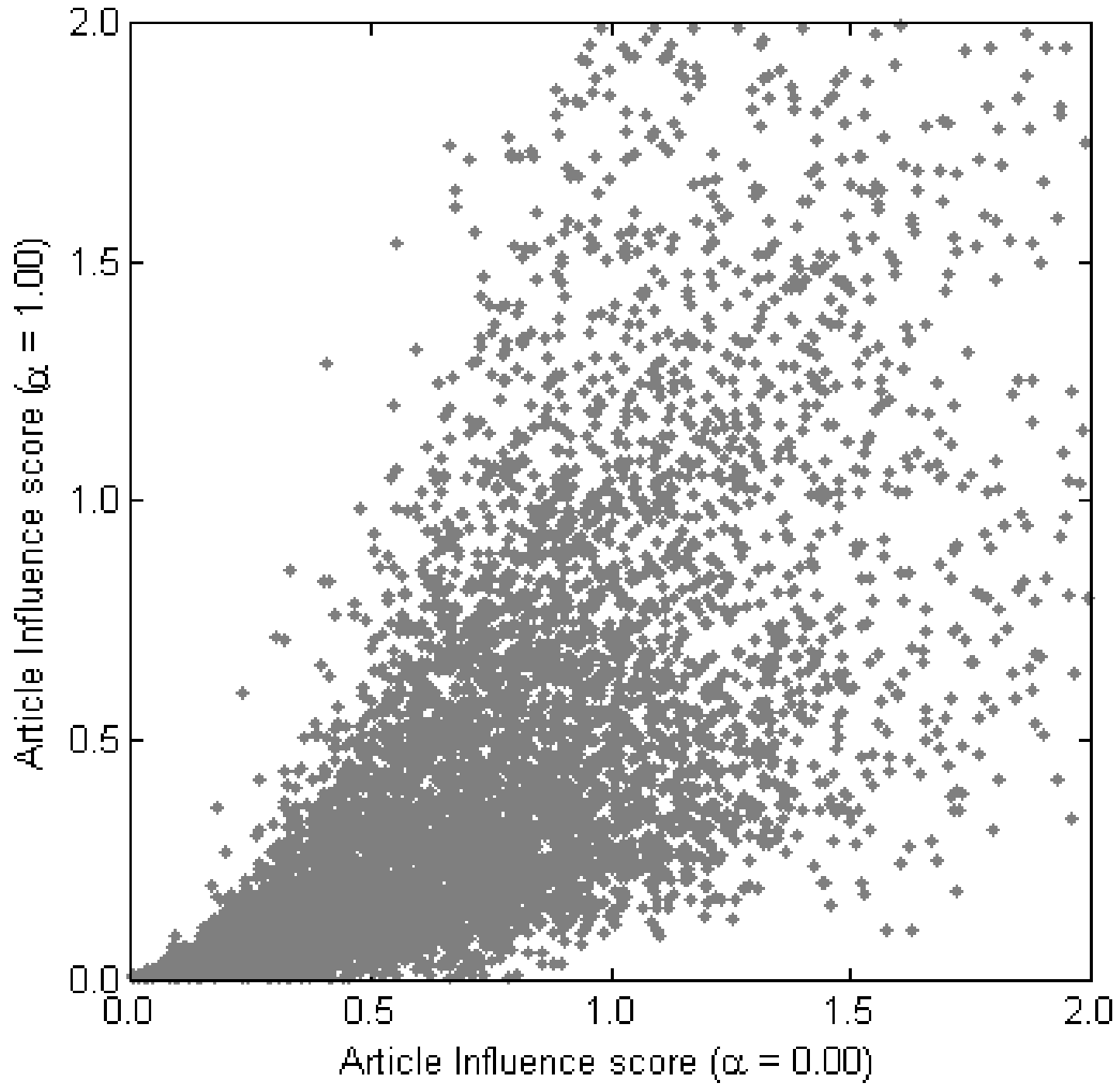

FIG. 3. Relation between the article influence indicator for $\alpha$ equal to 0 and the article influence indicator for $\alpha$ equal to 1. Out of the 6708 journals, only the 6252 journals with values below 2 for both indicators are shown.

Our empirical results show that the differences between indicators of journal performance are far from negligible. The results also show that the Eigenfactor parameter $\alpha$ has a quite large effect on the behavior of the article influence indicator (see especially Figure 3). Hence, based on our empirical analysis, it can be concluded that the study of different indicators is not merely of theoretical interest but also has a substantial practical relevance. We note that some papers (Davis, 2008; Fersht, 2009; Leydesdorff, 2009) report strong relations between certain indicators, which may seem to contradict our results. However, these papers either focus on indicators of total performance, for which it is not surprising to find strong relations (West et al., 2009), or they rely heavily on Pearson correlation scores. As shown in our analysis, high Pearson correlation scores may be somewhat misleading and should be interpreted with special care.

## Other indicators related to Eigenfactor

For completeness, in this section we briefly consider two other indicators of journal performance that are related to the Eigenfactor indicator. These indicators are the weighted PageRank indicator proposed by Bollen et al. (2006) and the SCImago Journal Rank indicator discussed by González-Pereira et al. (2009). SCImago Journal Rank scores of a large number of journals, calculated based on Scopus data, can be found at www.scimagojr.com. SCImago Journal Rank scores are also reported in the Scopus database. Like the Eigenfactor indicator, the weighted PageRank indicator and the SCImago Journal Rank indicator belong to the family of PageRank-inspired indicators.

Weighted PageRank scores are obtained by solving the following system of linear equations for $\beta \in [0, 1]$ and $\gamma = 0$:

$$r_i = \beta \sum_j \frac{r_j c_{ji}}{s_j} + \gamma \frac{a_{i1}}{\sum_j a_{j1}} + (1 - \beta - \gamma)\frac{1}{n} \qquad \text{for } i = 1, \ldots, n \tag{20}$$

$$\sum_i r_i = 1. \tag{21}$$

The weighted PageRank indicator is size dependent and measures a journal's total performance. SCImago Journal Rank scores are obtained in two steps. In the first step, the above system of linear equations is solved for $\beta$, $\gamma \in [0, 1]$ and $\beta + \gamma \leq 1$.[9] By default, $\beta$ and $\gamma$ are set equal to 0.9 and 0.0999, respectively.[10] In the second step, for each journal $i$ the SCImago Journal Rank score is calculated by dividing $r_i$ by $a_{i1}$. Since the SCImago Journal Rank indicator incorporates a normalization for the number of articles published in a journal, the indicator measures a journal's average performance per article. We note that calculating SCImago Journal Rank scores (or weighted PageRank scores) for $\beta + \gamma < 1$ has the effect that smaller journals are favored over larger ones. This seems an undesirable effect, and we therefore recommend choosing $\beta$ and $\gamma$ in such a way that $\beta + \gamma = 1$.

Although the weighted PageRank indicator and the SCImago Journal Rank indicator seem quite similar to the Eigenfactor indicator, there is a subtle but important difference. Choosing $\beta$ and $\gamma$ in such a way that $\beta + \gamma = 1$ and solving Equations 20 and 21 does not yield Eigenfactor scores. This is due to Equation 11 in the calculation of Eigenfactor scores. For this equation, there is no corresponding equation in the calculation of weighted PageRank scores or SCImago Journal Rank scores. A consequence of this observation is that the behavior of the weighted PageRank indicator and the SCImago Journal Rank indicator can be quite different from the behavior of the Eigenfactor indicator. For $\beta = 1$ and $\gamma = 0$, a result similar to Theorem 2 can be proven, which means that the weighted PageRank indicator and the SCImago Journal Rank indicator reduce to the influence per publication indicator of Pinski and Narin (1976). However, for $\beta = 0$ and $\gamma = 1$, there is no result similar to Theorem 1. This means that there is no relation between the SCImago Journal Rank indicator and the audience factor of Zitt and Small (2008). In fact, the SCImago Journal Rank indicator becomes quite meaningless for $\beta = 0$ and $\gamma = 1$. The indicator simply has the same value for all journals.

## Conclusions

In a recent report in which research assessment practices based on citation data are critically discussed, it is stated that the "assumptions behind (the Eigenfactor indicator) are not easy for most people to discern" and that the "complexity (of the Eigenfactor indicator) can be dangerous because the final results are harder to understand" (Adler, Ewing, & Taylor, 2009, p. 12). These are valid concerns that require serious attention. In this paper, we have addressed these concerns by providing some new insights into the mechanics of the Eigenfactor indicator. Most importantly, we have shown the close relation of the Eigenfactor indicator with the audience factor (Zitt & Small, 2008) and the influence weight indicator (Pinski & Narin, 1976). We have also introduced two properties that bibliometric indicators of journal performance may or may not have. These are the properties of insensitivity to field differences and insensitivity to insignificant journals. Based on the relation between the Eigenfactor indicator, the audience factor, and the influence weight indicator, we have pointed out that the Eigenfactor indicator (or, more precisely, its normalized variant, the article influence indicator) implements a trade-off between these two properties. In this way, we have also been able to give a concrete interpretation to the parameter of the Eigenfactor indicator. The empirical analysis that we have presented has shown that in practice the differences between various indicators of journal performance are quite substantial. This further illustrates the importance of having a good understanding of the properties of different indicators.

[9] The SCImago Journal Rank indicator uses different citation windows in the numerator and the denominator of the first term in Equation 20. For simplicity, we ignore this issue.

[10] In an earlier version of the SCImago Journal Rank indicator, the default values of $\beta$ and $\gamma$ were 0.85 and 0.1, respectively.

## Acknowledgment

We would like to thank three referees for their useful comments on an earlier draft of this paper.

## Appendix A: The property of insensitivity to field differences

In this appendix, we introduce the property of insensitivity to field differences. We study for different indicators of journal performance whether they have this property or not.

We first introduce the mathematical notation that we use. Suppose two fields can be distinguished, field 1 and field 2. There are $n_1$ journals in field 1, $n_2$ journals in field 2, and $n = n_1 + n_2$ journals in total. $J_1 = \{1, \ldots, n_1\}$ denotes the set of all journals in field 1, $J_2 = \{n_1 + 1, \ldots, n\}$ denotes the set of all journals in field 2, and $J = J_1 \cup J_2$ denotes the set of all journals. **A** denotes a positive matrix of size $n \times 2$. Elements $a_{i1}$ and $a_{i2}$ of **A** denote the number of articles published in journal $i$ in, respectively, periods $T_1$ and $T_2$. **A** satisfies

$$\sum_{i \in J_1} a_{i1} = \sum_{i \in J_2} a_{i1} . \tag{22}$$

Hence, the number of articles published in field 1 in period $T_1$ equals the number of articles published in field 2 in period $T_1$. **C** denotes the journal citation matrix. This is a non-negative matrix of size $n \times n$. Element $c_{ij}$ of **C** denotes the number of citations from articles published in journal $i$ in period $T_2$ to articles published in journal $j$ in period $T_1$.

Using the above mathematical notation, the property of insensitivity to field differences can be formally defined as follows.

**Property 1.** Let $f$ denote an indicator of a journal's average performance per article. $f$ is said to be *insensitive to field differences* if and only if

$$(1-\delta)\frac{\sum_{i \in J} a_{i1} f_i(\mathbf{A},\mathbf{C})}{\sum_{i \in J} a_{i1}} \leq \frac{\sum_{i \in J_k} a_{i1} f_i(\mathbf{A},\mathbf{C})}{\sum_{i \in J_k} a_{i1}} \leq (1+\delta)\frac{\sum_{i \in J} a_{i1} f_i(\mathbf{A},\mathbf{C})}{\sum_{i \in J} a_{i1}} \tag{23}$$

for $k = 1, 2$, for all $n_1$, $n_2$, **A**, and **C**, and for all $\delta$ such that

$$\sum_{j \in J_k} c_{ij} \geq (1-\delta) s_i \tag{24}$$

for $k = 1, 2$ and for all $i \in J_k$.

Informally, the property of insensitivity to field differences has the following interpretation. Suppose that there are two equally-sized fields and that each journal gives away at most a fraction $\delta$ of its citations to journals that are not in its own field. An indicator of journal performance is then said to be insensitive to field differences if the average value of the indicator for each field separately deviates no more than a fraction $\delta$ from the average value of the indicator for both fields together. Hence, in the case of two fields without any between-fields citation traffic, the property of insensitivity to field differences implies that the average value of an indicator is the same for both fields.

It is easy to see that the impact factor is not insensitive to field differences. This is not surprising, since it is well known that impact factors of journals in different fields should not be directly compared with each other. The following theorem states that under a relatively mild assumption the audience factor of Zitt and Small (2008) is insensitive to field differences.

**Theorem 3.** Let the number of articles published in a journal in period $T_2$ be proportional to the number of articles published in a journal in period $T_1$, that is, let $a_{i2}$ be proportional to $a_{i1}$. The audience factor then is insensitive to field differences.

**Proof.** We use the mathematical notation introduced at the beginning of this appendix. Let $a_{i2}$ be proportional to $a_{i1}$, that is, let there exist a constant $\eta > 0$ such that $a_{i2} = \eta a_{i1}$ for all $i \in J$. Let $\delta$ be chosen in such a way that Equation 24 is satisfied for $k = 1, 2$ and for all $i \in J_k$. It follows from Equations 3 and 4 that

$$\frac{\sum_{i \in J_k} a_{i1} \mathrm{AF}_i}{\sum_{i \in J_k} a_{i1}} = \frac{m_\mathrm{S} \sum_{i \in J_k} \sum_{j \in J} a_{j2} c_{ji} / s_j}{\sum_{i \in J_k} a_{i1}}. \tag{25}$$

Equation 24 implies that

$$(1 - \delta) s_j \leq \sum_{i \in J_k} c_{ji} \leq s_j \tag{26}$$

if $j \in J_k$ and that

$$0 \leq \sum_{i \in J_k} c_{ji} \leq \delta s_j \tag{27}$$

if $j \in J \setminus J_k$. Combining Equations 25, 26, and 27 yields

$$\frac{(1 - \delta) m_\mathrm{S} \sum_{j \in J_k} a_{j2}}{\sum_{i \in J_k} a_{i1}} \leq \frac{\sum_{i \in J_k} a_{i1} \mathrm{AF}_i}{\sum_{i \in J_k} a_{i1}} \leq \frac{m_\mathrm{S} \sum_{j \in J_k} a_{j2} + \delta m_\mathrm{S} \sum_{j \in J \setminus J_k} a_{j2}}{\sum_{i \in J_k} a_{i1}}. \tag{28}$$

Taking into account that $a_{i2}$ is proportional to $a_{i1}$, it follows from Equations 22 and 28 that

$$(1 - \delta) \eta m_\mathrm{S} \leq \frac{\sum_{i \in J_k} a_{i1} \mathrm{AF}_i}{\sum_{i \in J_k} a_{i1}} \leq (1 + \delta) \eta m_\mathrm{S}. \tag{29}$$

It can further be seen that

$$\frac{\sum_{i \in J} a_{i1} \mathrm{AF}_i}{\sum_{i \in J} a_{i1}} = \eta m_\mathrm{S}. \tag{30}$$

Equations 29 and 30 imply that

$$(1-\delta)\frac{\sum_{i\in J} a_{i1}\mathrm{AF}_i}{\sum_{i\in J} a_{i1}} \leq \frac{\sum_{i\in J_k} a_{i1}\mathrm{AF}_i}{\sum_{i\in J_k} a_{i1}} \leq (1+\delta)\frac{\sum_{i\in J} a_{i1}\mathrm{AF}_i}{\sum_{i\in J} a_{i1}}. \tag{31}$$

Hence, Equation 23 is satisfied, which means that the audience factor has Property 1. This completes the proof of the theorem.

The influence per publication indicator of Pinski and Narin (1976) is not insensitive to field differences. This is stated in the following theorem.

**Theorem 4.** The influence per publication indicator is not insensitive to field differences.

**Proof.** We prove the theorem by means of a counterexample. We use the mathematical notation introduced at the beginning of this appendix. Let $n_1 = n_2 = 1$, let

$$\mathbf{A} = \begin{bmatrix} 100 & 100 \\ 100 & 100 \end{bmatrix}, \tag{32}$$

$$\mathbf{C} = \begin{bmatrix} 999 & 1 \\ 3 & 997 \end{bmatrix}, \tag{33}$$

and let $\delta = 0.003$. Equation 24 is then satisfied for $k = 1, 2$ and for all $i \in J_k$. It follows from Equations 6, 7, and 8 that

$$\frac{\sum_{i\in J_1} a_{i1}\mathrm{IPP}_i}{\sum_{i\in J_1} a_{i1}} = 15 > 10.03 = (1+\delta)\frac{\sum_{i\in J} a_{i1}\mathrm{IPP}_i}{\sum_{i\in J} a_{i1}}. \tag{34}$$

Hence, Equation 23 is not satisfied, which means that the influence per publication indicator does not have Property 1. This completes the proof of the theorem.

The proof of Theorem 4 illustrates an important problem of the influence per publication indicator. When there are two fields and there is almost no citation traffic between the fields, influence per publication scores become extremely sensitive to the exact number of times one field cites the other (see also West et al., 2008, p. 3). In other words, the influence per publication indicator becomes unstable when the journal citation matrix is almost reducible. This problem is also discussed by Serrano (2004). For a thorough mathematical treatment of this issue for PageRank-inspired indicators (of which the influence per publication indicator can be seen as a limit case), we refer to Langville and Meyer (2006, Section 6.1).

## Appendix B: Best performing journals

TABLE 4. The 20 best performing journals according to four indicators of journal performance.

| Journal | IF | Journal | AI(0.00) |
|---|---|---|---|
| Annual Review of Immunology | 37.7 | Reviews of Modern Physics | 16.5 |
| CA-A Cancer Journal for Clinicians | 37.4 | CA-A Cancer Journal for Clinicians | 14.7 |
| Reviews of Modern Physics | 31.8 | New England Journal of Medicine | 12.0 |
| New England Journal of Medicine | 29.5 | Annual Review of Immunology | 10.2 |
| Physiological Reviews | 29.3 | Materials Science & Engineering R-Reports | 9.6 |
| Annual Review of Biochemistry | 27.7 | Physiological Reviews | 9.3 |
| Nature Reviews Cancer | 27.0 | Chemical Reviews | 9.0 |
| Nature Reviews Immunology | 26.5 | Annual Review of Biochemistry | 8.4 |
| Nature Reviews Molecular Cell Biology | 26.5 | Nature Reviews Cancer | 8.0 |
| Annual Review of Neuroscience | 24.9 | Nature Materials | 8.0 |
| Chemical Reviews | 23.9 | Progress in Materials Science | 7.9 |
| Cell | 22.3 | Progress in Polymer Science | 7.8 |
| Annual Review of Cell and Developmental Biology | 21.1 | Nature | 7.6 |
| Nature | 21.1 | JAMA-Journal of the American Medical Association | 7.5 |
| Nature Reviews Neuroscience | 20.9 | Annual Review of Neuroscience | 7.5 |
| Nature Immunology | 20.5 | Nature Reviews Molecular Cell Biology | 7.5 |
| Nature Medicine | 20.3 | Science | 7.3 |
| Science | 20.0 | Nature Reviews Immunology | 7.3 |
| Nature Genetics | 18.7 | Surface Science Reports | 7.0 |
| Endocrine Reviews | 18.4 | Annual Review of Fluid Mechanics | 6.5 |

| Journal | AI(0.85) | Journal | AI(1.00) |
|---|---|---|---|
| Annual Review of Immunology | 22.0 | Annual Review of Immunology | 33.3 |
| Reviews of Modern Physics | 20.5 | Annual Review of Biochemistry | 27.4 |
| Annual Review of Biochemistry | 18.2 | Cell | 26.5 |
| Nature Reviews Molecular Cell Biology | 16.9 | Nature Reviews Molecular Cell Biology | 26.1 |
| Cell | 16.8 | Annual Review of Neuroscience | 24.8 |
| Annual Review of Neuroscience | 16.7 | Annual Review of Cell and Developmental Biology | 22.5 |
| CA-A Cancer Journal for Clinicians | 16.0 | Nature Reviews Immunology | 21.8 |
| Nature Reviews Immunology | 14.7 | Nature Immunology | 20.6 |
| New England Journal of Medicine | 14.6 | Nature Genetics | 20.0 |
| Nature | 14.4 | Annual Review of Astronomy and Astrophysics | 19.8 |
| Annual Review of Cell and Developmental Biology | 14.3 | Nature | 19.7 |
| Physiological Reviews | 13.7 | Reviews of Modern Physics | 19.4 |
| Nature Reviews Cancer | 13.6 | Nature Reviews Cancer | 19.0 |
| Nature Genetics | 13.4 | Physiological Reviews | 18.8 |
| Science | 13.3 | CA-A Cancer Journal for Clinicians | 18.0 |
| Nature Immunology | 13.1 | Science | 17.3 |
| Quarterly Journal of Economics | 12.6 | Nature Reviews Neuroscience | 17.0 |
| Nature Reviews Neuroscience | 11.8 | New England Journal of Medicine | 16.9 |
| Nature Medicine | 10.8 | Nature Cell Biology | 15.3 |
| Nature Materials | 10.4 | Immunity | 15.2 |